# Development of a Device for Remote Monitoring of Heart Rate and Body Temperature

Mohammad Ashekur Rahman[1], Atanu Barai[2], Md. Asadul Islam[3], M.M.A Hashem[4]
Department of Computer Science and Engineering
Khulna University of Engineering & Technology (KUET)
Khulna 9203, Bangladesh
ashekcse@gmail.com[1], atanu.bct@gmail.com[2], asad_kuet@yahoo.com[3], mma_hashem@hotmail.com[4]

***Abstract*—We present a new integrated, portable device to provide a convenient solution for remote monitoring heart rate at the fingertip and body temperature using Ethernet technology and widely spreading internet. Now a day, heart related disease is rising. Most of the times in these cases, patients may not realize their actual conditions and even it is a common fact that there are no doctors by their side, especially in rural areas, but now a day's most of the diseases are curable if detected in time. We have tried to make a system which may give information about one's physical condition and help him/her to detect these deadly but curable diseases. The system gives information of heart rate and body temperature simultaneously acquired on the portable side in real-time and transmits results to web. In this system, the condition of heart and body temperature can be monitored from remote places. Eventually, this device provides a low-cost, easily accessible human health monitor solution bridging the gaps between patients and doctors.***

***Keywords*—Body Temperature, Ethernet, Heart Rate, Infrared Transmitter, Infrared Receiver, Microcontroller**

## I. INTRODUCTION

Heart rate means the number of heartbeats per unit of time, usually expressed as beats per minute (bpm). Human's heart pounds to pump oxygen-rich blood to muscles and to carry cell waste products away from tissues [1]. Heart rate can vary according to the demand of muscles to absorb oxygen and excrete carbon dioxide changes, such as during exercise or sleep [2]-[4]. It also varies significantly between individuals based on fitness, age and genetics. That means heart must beat faster to deliver more oxygen-rich blood. During exercise routines, the heart rate gives a strong indication of how effective that routine is improving health. Normal heart rate of a resting person is about 70 bpm for adult males and 75 bpm for adult females. A heart rate monitor is simply a device that takes a sample of heartbeats and computes the beats per minute so that the information can easily track heart condition [5].

Medical professionals use heart rate for tracking of patient's physical conditions. Individuals, such as athletes, who are interested in monitoring their heart rate to gain maximum efficiency from their training, also use it.

Body temperature means measurement of the body's ability to generate and get rid of heat. It is one of chief indicators of normal functioning and health. The nature of the human body is to keep its temperature within a narrow, safe range in spite of large variations in temperatures outside the body. Normal human body temperature depends upon the place in the body, from which the measurement is made, and the time and level of activity of the person. The typical body temperature is 37.0 °C ± 0.4°C (98.6 ° F ± 0.7°F).

When one is too hot, the blood vessels in his/her skin expand (dilate) to carry the excess heat to his/her skin's surface. One may begin to sweat, and as the sweat evaporates, it helps to cool his/her body. When one is too cold, his/her blood vessels narrow (contract) so that blood flow to his/her skin is reduced to conserve body heat. One may start shivering, which is an involuntary, rapid contraction of the muscles. This extra muscle activity helps generate more heat. Under normal conditions, this keeps one's body temperature within a narrow, safe range.

The remote heart rate and temperature monitoring system gives information of heart rate and body temperature simultaneously and sends results to the web server. From the web, anyone can monitor the physical status of the patient. Thus in this system the condition of the body can be monitored from remote places.

The later part of this paper shows the problem statement, background information, system overview, methodology, actual hardware test result of our device.

## II. BACKGROUND INFORMATION

### *A. Problem Definition*

One of the increasing popular public concerns is human health. Anything else becomes meaningless if one gets sick or dead. For this reason, people spend a lot of money to keep sound health. Unfortunately, people always find that it is too late to receive serious medical care when things are non-invertible. If early actions can be taken in time then lots of patients can be cured. However, access to many medical equipment is inconvenient and expensive. Heart rate and body temperature are the most vital ones among the most notable indexes of the human health, and they have the advantage of easy access. Moreover, unlike the X-ray, the measurement of heart rate and body temperature has no effect on human health itself.

There are some devices in the current market which can provide raw medical measurement data to patients and doctors, but the patients may not interpret the medical measurement into meaningful diagnosis since they have little medical background. On the other hand, if raw medical data is delivered to the doctor, it kills much time and may cause trouble, but in emergencies time can never be wasted. It is tough to share data over a large area within a short period. Most of the products available in the current market have these major drawbacks with limitation in flexibility and portability.

*B. Existing Approaches*

Heart rate and body temperature are the most vital ones among the most important indexes of the human health, and they have the advantage of easy access. There are some existing approaches for monitoring heart rate. The current technology for measuring heart rate consists of optical and electrical methods. The electrical method provides a bulky strap around one's chest [6]. The optical method requires no such strap and can be used more conveniently than the electrical method. Low cost heart rate measurement device was developed using optical technology. In optical method, there is an approach using powerful LED and Light Dependent Register (LDR) to sense pulses [7]. Here, the pulses are amplified by an amplifier circuit and filtered by a band pass filter. After that the amplified and filtered pulse signals are sent to the microcontroller. The microcontroller receives the pulse signals as analog signals and uses a standard voltage to check if the pulse signals are valid or not. Then the heart rate is counted by that microcontroller and displayed the result in a LCD display. There is also an another approach using infrared Tx and Rx which also senses pulses, amplifies the pulses and filters pulses by a low pass filter. Then the pulse signal is sent to a microcontroller [8], [9]. The microcontroller receives the pulses as analog signals and checks the signals with a standard voltage and does the same thing as the above approach. Both the approach use analog signal to measure heart rate, but analog signal of pulse vary from person to person. These approaches cannot calibrate analog signal of pulses for each person and use a standard voltage to check each pulse signal. That why these approaches give inaccurate result in many cases. There is also an approach which uses infrared technology to measure heart rate and use analog temperature sensor to measure body temperature. This approach can calibrate analog signal of pulses for each person. It uses wireless terminal to send data and wireless receiver to receive the data. The data are then sent to the computer using serial port. Data is then sent to the web server using the internet and can be viewed from anywhere in the web browser. The main drawback of this approach is that it needs a computer to send data to the web server through the internet.

Based on these circumstances, we thought to make a device which enables people to do the measurements themselves and share it to whoever it may concern automatically without computer over both local network and internet. The goal of our project is to design a remote heart rate and body temperature monitoring device which is inexpensive, accurate, durable, and affordable and user friendly using Infrared and Ethernet technology. We have incorporated the infrared technology using a standard 5mm Infrared LED (IRTx) and photo transistor (IRRx) and measured body temperature using LM35. The IRTx and IRRx have been used to measure the heart rate from any finger. A microcontroller has been programmed to count the pulse rate and body temperature. The heart rate and body temperature is digitally displayed on a LCD display controlled by the same microcontroller that counts the pulse rate and body temperature and transmits to the web server using Ethernet technology. The significance of the heart rate and body temperature monitor is that it provides an inexpensive and accurate means of measuring one's health condition at his/her convenience.

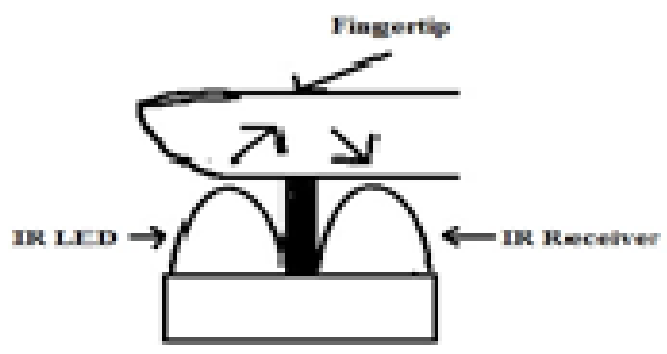

Figure 1. Finger Placement

## III. SYSTEM OVERVIEW

There are many constraints in designing a portable heart rate and body temperature monitoring system which will remotely monitor health condition. First, the technology used to measure the pulse has been determined. As the basis of our design is to construct an efficient and affordable remote heart rate and temperature monitor, an infrared Tx and a Rx have been used to measure the pulse by measuring the change in blood flow through one of the fingers. A noise filter has been designed to filter out any unwanted noise and interference, due to small movement of finger causes excess noise. A temperature sensor has been used to measure body temperature. For streaming live health condition, we have developed a system that not only measures heart rate and body temperature accurately but also transmits data simultaneously to the web server. A microcontroller has been programmed to count the pulse rate, monitor signal of pulse and measure the body temperature and control a display to show the measured data and effectively send data to the web server through Ethernet and show the pulse rate, temperature and pulse waveform on a web browser. The device operates in 5 volts and can facilitate battery operation. The battery life is reasonably high under normal use with this nominal power consumption. The final product is packaged in a small, lightweight, and durable package. To achieve the final objective the following sub objectives has been carried out:

*A. Heart Rate*

*A.1. Pulse Detection*

Optical sensors have been used to measure the alteration in blood volume at fingertip with each heart bit. The sensor unit consists of an IRTx and an IRRx, placed side by side. The finger placement is shown in Fig. 1. The IRTx transmits an infrared light into the fingertip (placed over the sensor unit), and the IRRx senses the portion of the light that is reflected back. The intensity of reflected light depends upon the blood volume inside the fingertip. So, each heart bit slightly alters the amount of reflected infrared light that can be detected by the IRRx. With a proper signal conditioning this little change in the amplitude of the reflected light can be converted into pulse.

*A.2. Signal Conditioning and Amplifying*

The signal conditioning circuit consists of two identical active low pass filters with a cut-off frequency of about 2.34 Hz. This means the maximum measurable heart rate is about 140 bpm. LM358 dual OpAmp chip has been used in this circuit which operates at a single power supply. The filter blocks any higher frequency noises present in the signal. The two stage filter and amplifier converts weak signal coming from the photo sensor unit into a pulse. An LED connected at the output blinks every time indicating a heart bit is detected. The pulse generated from the signal conditioner and amplifier is fed to the digital IO pin P5 and analogue input pin P19 of mbed NXP LPC1768 [10].

*B. Body Temperature*

The LM35 temperature sensor has been used to measure the body temperature. The output pin of the sensor is connected to an analogue input pin P18 of the microcontroller.

*C. Display*

The measured heart rate and the body temperature have been displayed on a LCD display. A microcontroller has been programmed for this reason. As improper placement of the finger would create noise and give improper data, a flashing segment on the LCD display has been used to help the user to identify when heart rate and body temperature measurement have started.

*D. Ethernet Host*

An Ethernet host of the microcontroller has been used to send the measured data through the internet to the web server. For this purpose, the microcontroller has been connected to a RJ45 jack.

*E. Web Server*

The data received from the device is shown in the web browser, in real-time. An ECG graph is also plotted in the browser which demonstrates the condition of the heart bit.

*F. Accuracy*

The heart rate and body temperature measured from our device give accurate result with no more than +/- 1% error. The ECG gives accurate result with no more than +/- 10% error. The ECG is not quite as tight as the ECG monitors (+/- 1%), but cost of the measurement device can be reduced by using our design. More accuracy can be gained by taking more samples.

*G. Power*

The device operates using a 5-volt battery source because of the small packaging being utilized by our design. Maximum power consumption of the device is about 0.70W. In sleep mode 0.64W, in power down mode 0.49W and in deep power down mode 0.29W.

*H. Durability*

The designed device operates in a temperature environment of 00C to 700C C. The device packaging has been designed to hold up under normal usage. The packaging is shock and but not water-resistant.

*I. Physical Packaging*

The design of this device is small and lightweight which makes it easily portable. It is approximately the size of one's palm for easy usage. The final packaged dimensions may be no larger than a 3.5" x 2" x 1" (H x W x D).

*J. Cost*

At the development phase, the cost for the components of the device is around \$65. This cost can be reduced by 40% if a custom microcontroller is designed for mass production, which has the features only required for this design.

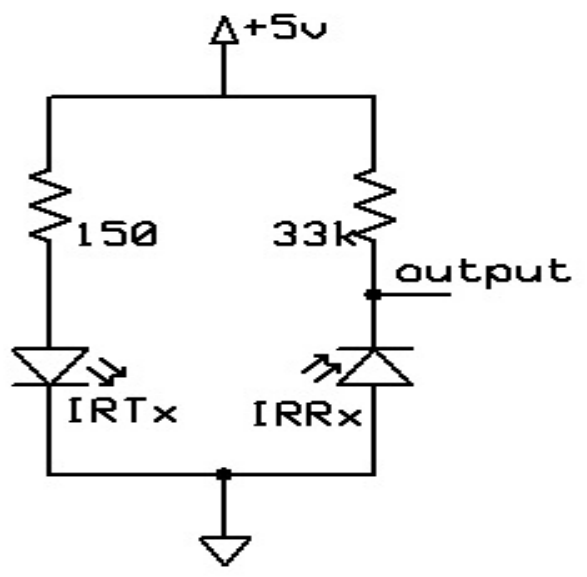

Figure 2. Pulse Detection Circuit

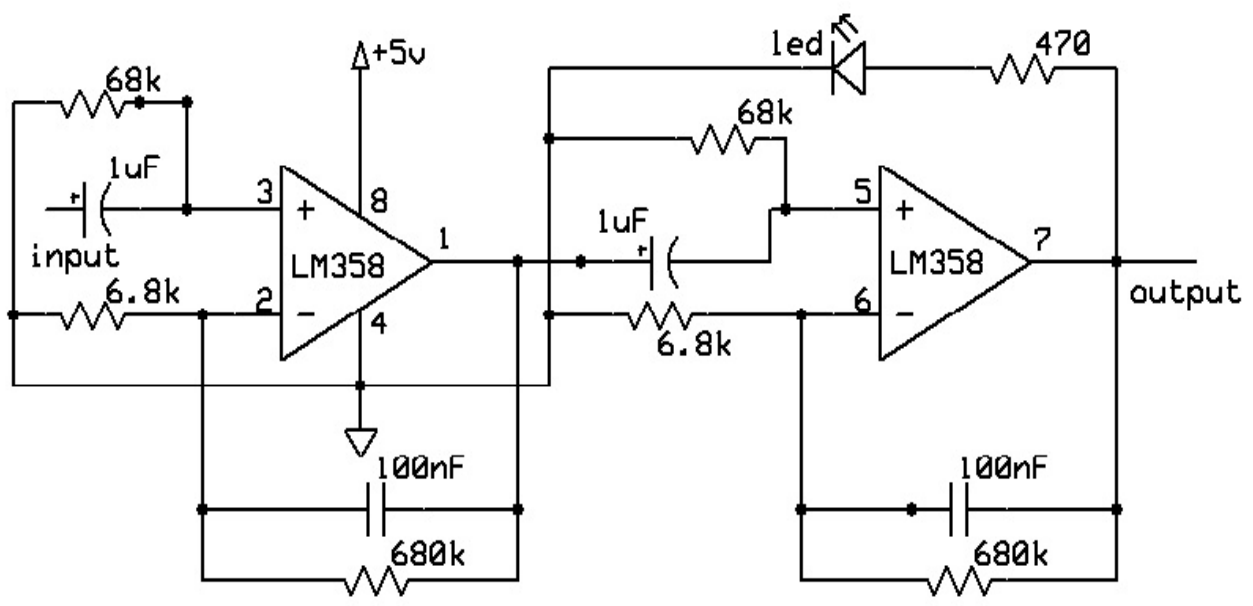

Figure 3. Signal Filtering and Amplifying Circuit [8]

## IV. METHODOLOGY

*A. Infrared Transmitter and Receiver Circuit*

The haemoglobin molecules of blood absorb the infrared light [11]. Each time heart pumps, the volume of oxygen rich blood increases in the finger. As a result, the amount of oxyhaemoglobin molecules also increases in blood. Absorption of infrared light is also high and, reflection of infrared light is low. Then, each heart beat slightly alters the amount of reflected infrared light which can be detected by the IRRx. The circuit shown in Fig. 3 has been designed to do this. The more infrared light is received the less the voltage of the input point from the sensor part is produced. The IRRx picks an AC signal with some DC components. The DC components come up from non-pulsative tissues. Direct crosstalk between the IR transmitter and receiver is avoided though they are placed closely. A Resistor is connected to the Infrared receiver (IRRx) to reduce the current drawn by the detection system. If the intensity, of IR light is too high, then the reflected infrared light from the tissue will be sufficient enough to saturate the photo detecting diode all the time, and no signal will exist. So the value of the resistance connected in series with the Infrared transmitter (IRTx) is chosen to limit the current and hence the intensity of the transmitted infrared light.

The diagram in Fig. 2 shows the configuration of the IR emitter and IR receiver in relation to the finger. They are placed in such a way that the infrared light reflects from the finger to the to the IR receiver [3].

*B. Filtering and Amplification of pulse rate signal*

A small movement of finger causes high frequency noise. The pulse rate signal filtering is necessary to block any higher frequency noises present in the signal. The desired signal can be extracted from the noisy signal using a low pass filter.

The equation for cut off frequency of the low pass filter is given below.

$$Cutoff\ \ Frequency = 1/2\pi R_f C_f \tag{1}$$

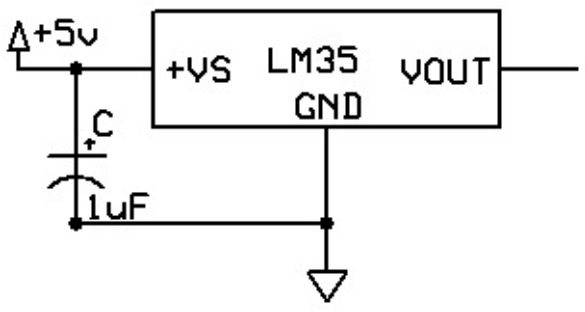

Figure 4. Temperature Sensor Circuit

The cut off frequency of the filter designed is 2.34 Hz. If not amplified, the signal found from the filter circuit is found having amplitude in mV level. The signal must be amplified for understanding and counting pulse rate by the microcontroller. A two stage signal filter and amplifier circuit using LM358 OpAmp can be designed for this. This OpAmp is operated with 5 volts. The designed circuit is shown in Fig. 4. The equation for gain of each stage of the low pass filter is given below.

$$Gain\ of\ Each\ Stage = 1 + R_f / C_f \quad (2)$$

In the designed circuit, total gain is 10201. Values of Rf and Ri are 680 KΩ and 6.8 KΩ. The 1uF capacitors, which are connected in series to the inputs of each filter blocks the undesired DC components of the signal. The two 1μF capacitors must be able to stand some reverse bias, so they should be nonpolarized.

*C. Body Temperature Sensing Circuit*

For body temperature measurement, LM35 sensor has been used. This sensor outputs analogue voltage according to temperature. A 1uF is connected across the Vin and ground of the sensor to reduce output drift due to noise in input voltage. Fig. 4 shows the circuit diagram for temperature sensor.

The temperature vs. voltage relation for basic Centigrade Temperature Sensor (+2˚C to +150˚C) is Vout = 0 mV + 10.0mV/˚C.

*D. Software Design for Microcontroller*

The microcontroller is used to calculate the number of heart bits per minute. It also measures the body temperature and sends the calculated values to the web server through Ethernet. The microcontroller drives a LCD display which shows the measured values locally. So any external LCD driver is not required. CMSIS RTOS for mbed has been used to take the real-time data. A thread continuously takes the analogue signal of heart condition and sends the values to the web server. If the signal voltage is above the threshold voltage, then it is considered as a heart bit is detected and counts up to 30 bits. A timer counts the time (t30) for 30 bits and calculates heart rate using (3).

$$Heart\ Rate(bpm) = 30 \times 60 / Measured\ Time\ (t_{30}) \quad (3)$$

The program also takes the analogue voltage output (Vt) of the temperature sensor and measures the body temperature using (4).

$$Body\ Temperature\ (^{\circ}F) = (V_t \times 3.3 \times 100) \times 9 \div 5 + 32 \quad (4)$$

The flowchart of the algorithm of the heart rate and body temperature is given in Fig. 5.

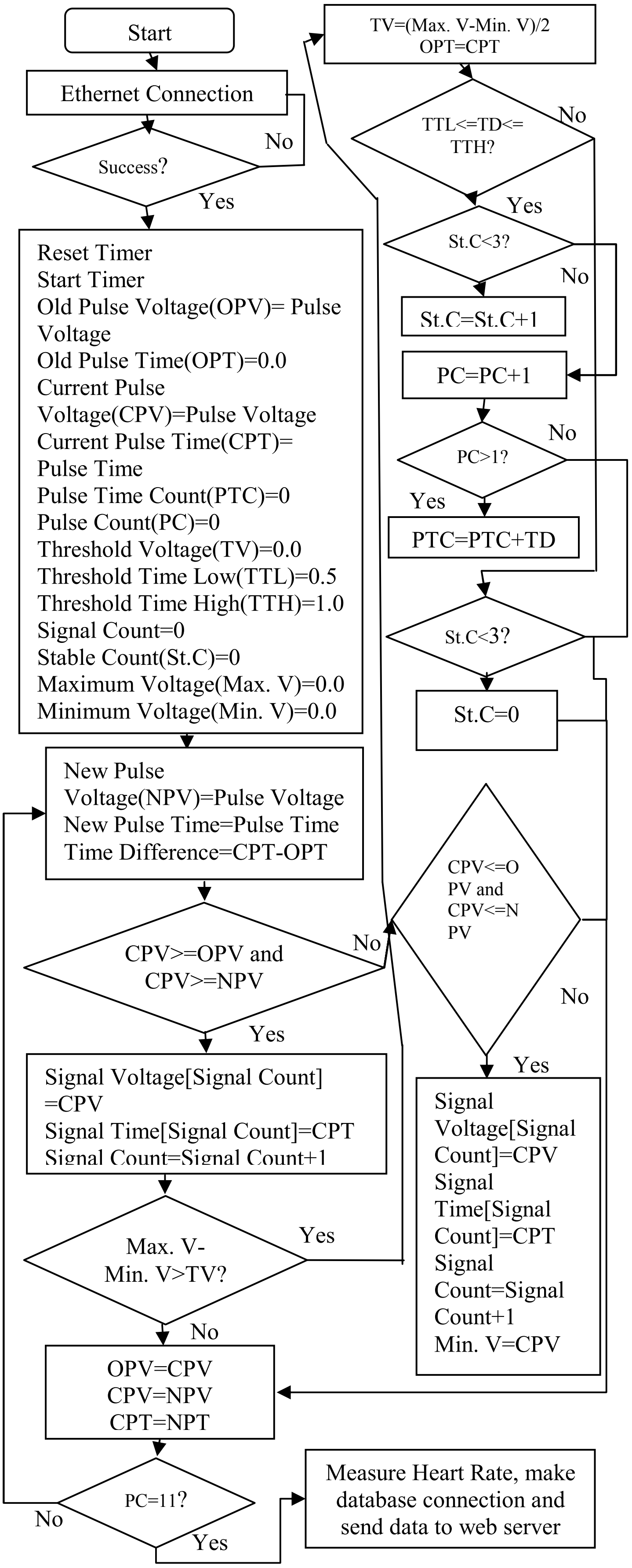

Figure 5. Algorithm for Measuring Heart Rate & Body Temperature

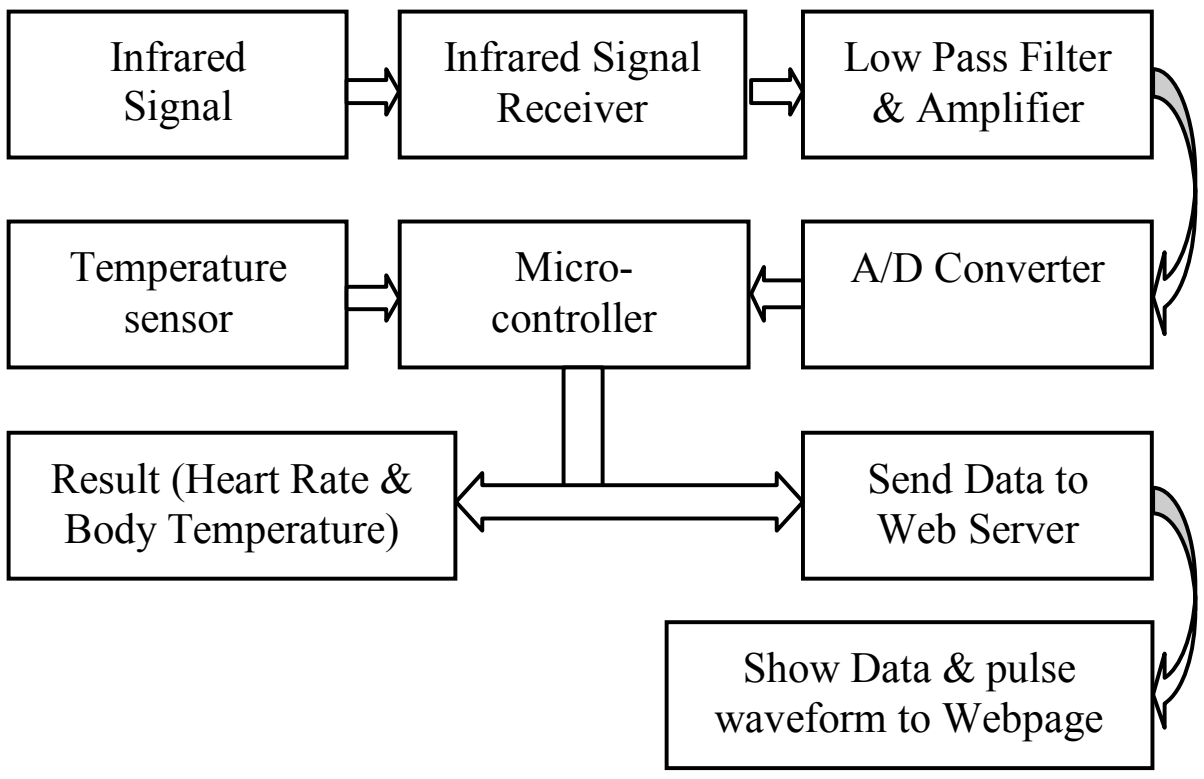

Figure 6. System Flowchart

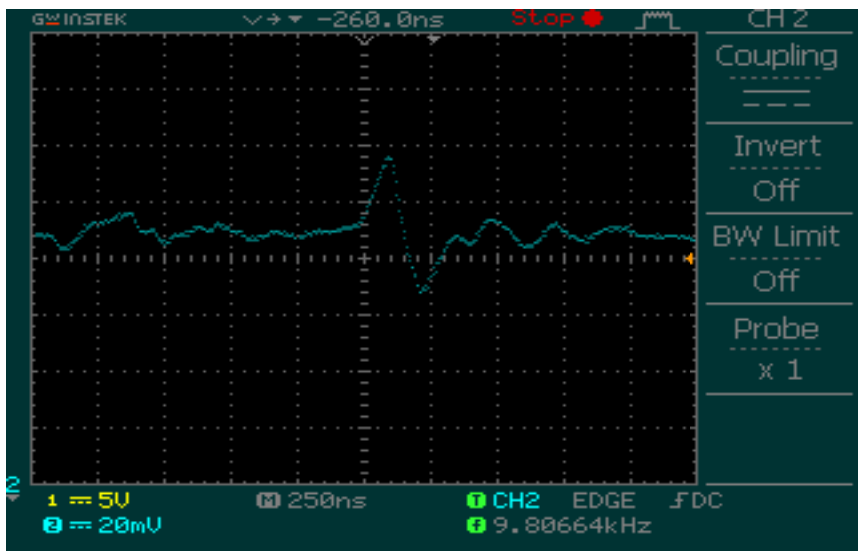

Figure 7. Heart Beat Signal

### *E. Complete Design Flow Chart*

As each stage of the design works successfully, they are tied together to make the whole system functional. The infrared light is transmitted and received using the circuit in Fig. 3. Next, using the low pass filter and amplifier circuit in Fig. 4 the signal is made computable by the microcontroller. The temperature sensor also outputs an analog signal. Once the signals are digitized by the microcontroller, they are computed and displayed using the microcontroller in a LCD display. The results are also sent to the web using Ethernet. In the web page, a graph is plotted which is similar to ECG of the patient. The calculated heart rate and body temperature is also shown in the web page. In Fig. 6 below, this process is illustrated.

## V. ACTUAL HARDWARE TEST RESULT

### *A. Optical Transmit and Receive Result*

The first step of the design, the infrared ray transmitter and receiver, is constructed and tested. The output of the receiver is connected to an oscilloscope to get the heart beat signal. Fig. 7 shows the heart beat signal obtained from the output of the receiver.

### *B. Signal Conditioning and Amplifying Result*

A low pass filter is used to filter the noise from the heart beat signal. Two stage amplifier is also used to amplify the heart beat signal. Fig. 8 and Fig. 9 shows the output obtained after the noisy heart signal is passed through the filter and amplifier. In Fig. 9, the second signal peak in a pulse should not be detected as a separate pulse. So a pick value is chosen such that when the amplitude of the signal is rises above the pick value, then it is detected by the microcontroller as a pulse.

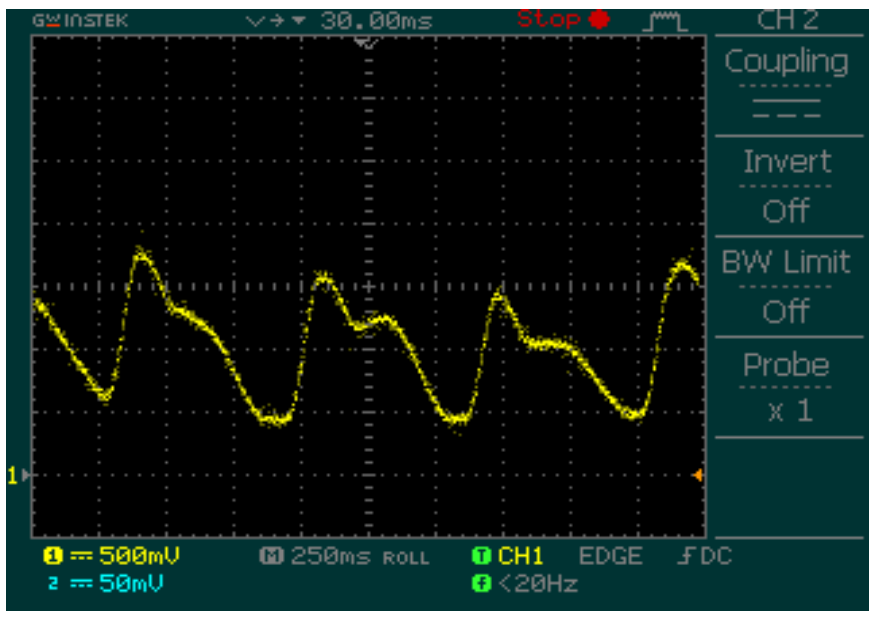

Figure 8. Output of First OpAmp

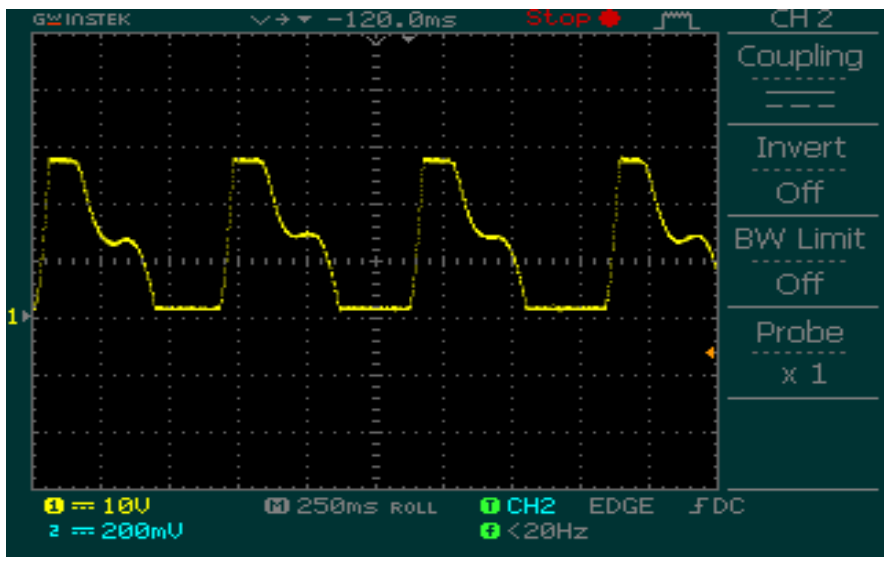

Figure 9. Output of Second OpAmp

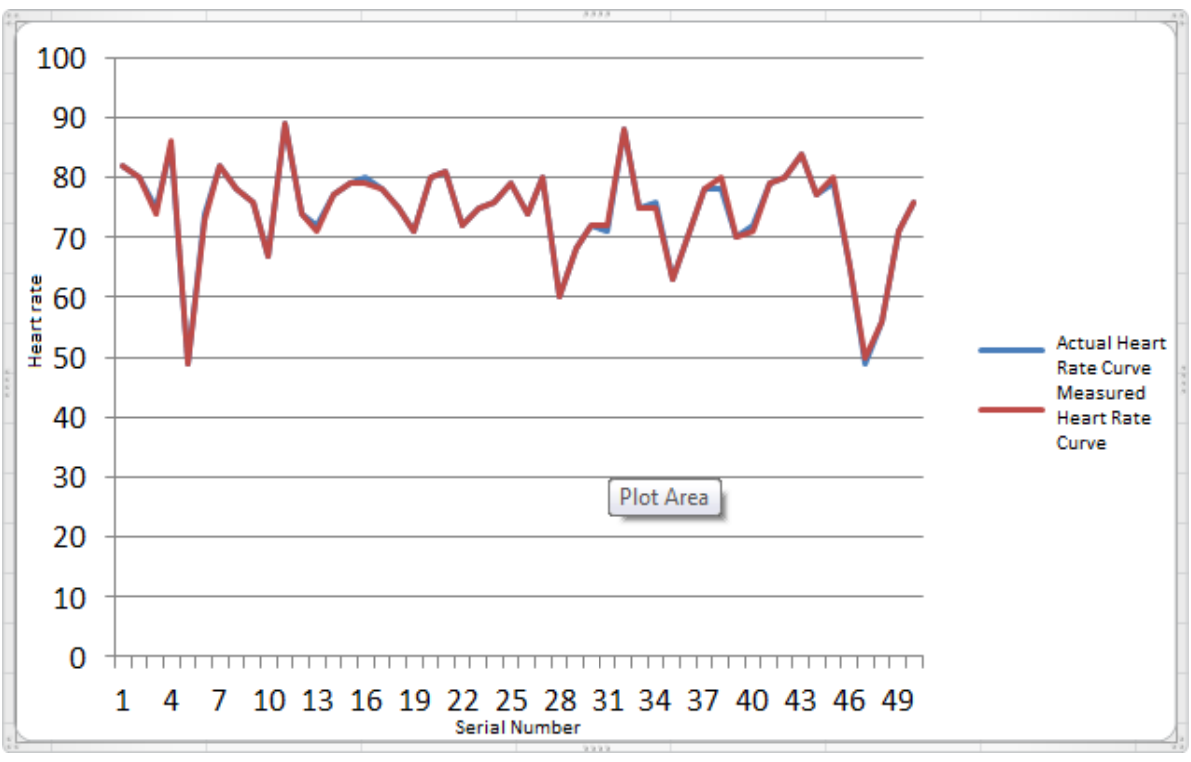

Figure 10. Curve for Actual vs. Measured Heart Rate

### *C. Microcontroller*

The microcontroller is programmed to get the analogue signal of the heart beat and send it to the web server. It is also programmed to count the time of 30 heart beats and measure the heart rate (bpm). It is also programmed to get the analogue signal from the temperature sensor and measure the body temperature (˚F). Then the microcontroller sends the data to the web server through Ethernet. The heart rate and the body temperature are shown on the LCD display. When the microcontroller is integrated into the entire design circuitry, it is able to count the number heart beats per minute and drive the LCD display to display the counted value.

### *D. Actual vs. Measured Heart Rate and Body Temperature*

The accuracy of the heart rate monitor is tested. Fig. 10 shows a dual curve where red curve shows the measured heart rate and blue curve shows the actual heart rate for some practical data. We see that there is remarkably small difference between the two curves in the dual curve which mirrors that the error rate is low. Fig. 11 shows a dual curve where red curve shows the measured body temperature and blue curve shows the actual body temperature for some practical data. We see that there is a small difference between the two curves in the dual curve which mirrors that the error rate is low.

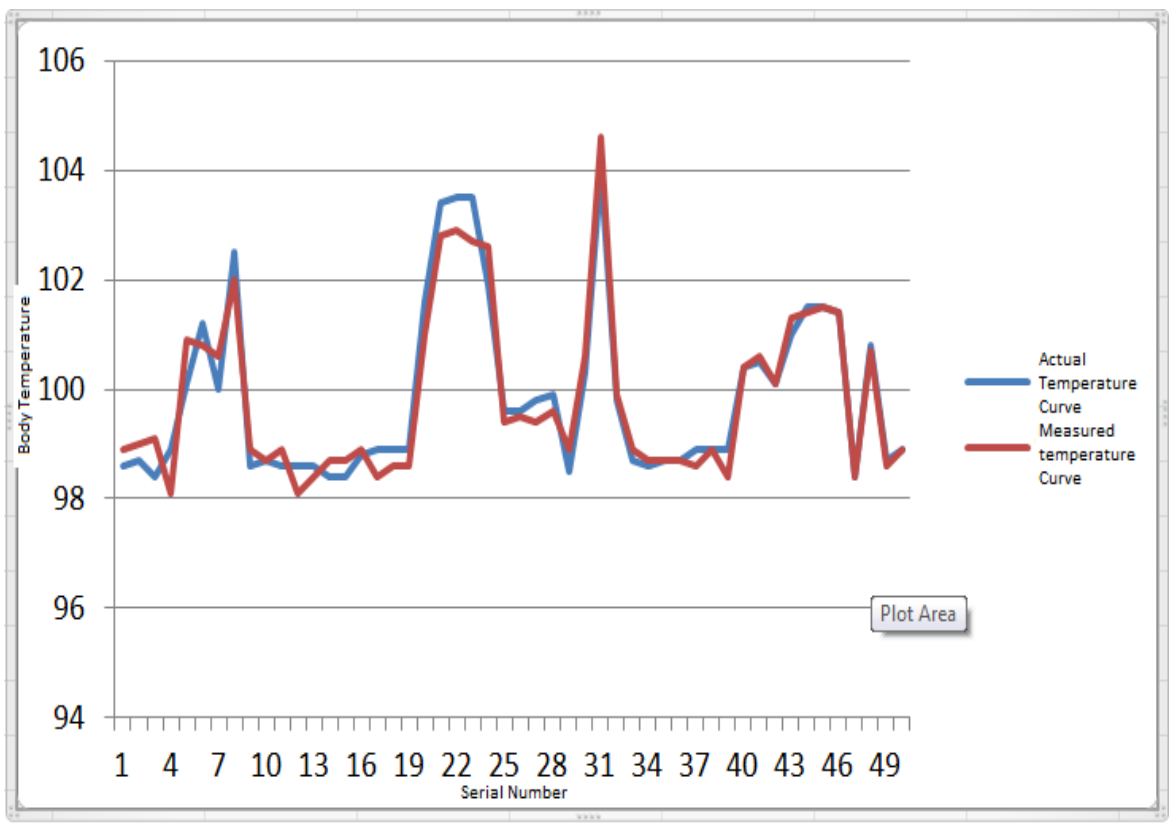

Figure 11. Curve for Actual vs. Measured Body Temperature

## VI. CONCLUSION

In this paper, we have shown a design of a new remote heart rate and body temperature monitoring device. The final result of our approach is a remote health condition measurement system with a flexible architecture that can be adopted in several different application fields. The system has been tested and valid for some bio signal such as heart rate and body temperature. The bio signals are measured in a real time with a higher accuracy but more cost effective than the old hand measuring system.

## ACKNOWLEDGMENT

The authors would like to thank Isfar Ahmed Sifat and all other teachers of CSE department of KUET and other authors who helped the authors providing guidelines to perform the work.